\documentclass[reprint,amsmath,amssymb,aip,eqsecnum,jcp]{revtex4-1}
\usepackage[dvips]{graphicx}
\usepackage{color}
\usepackage{amsmath}
\usepackage{amssymb}
\usepackage{pstricks,pst-grad}
\usepackage{amsbsy}
\usepackage{epic}
\usepackage[normalem]{ulem}

\usepackage[linktoc=all,unicode=true,pdfusetitle,  bookmarks=true,bookmarksnumbered=false,bookmarksopen=false, breaklinks=true,pdfborder={0 0 0},backref=false,colorlinks=true]{hyperref}
\hypersetup{linkcolor=blue,citecolor=red}

\usepackage[hyphenbreaks]{breakurl}

\usepackage{dcolumn}
\usepackage{bm}

\def\bal#1\eal{\begin{align}#1\end{align}}
\newcommand\beq{\begin{equation}}
\newcommand\eeq{\end{equation}}
\newcommand\beqa{\begin{eqnarray}}
\newcommand\eeqa{\end{eqnarray}}
\newcommand{\nn}{\nonumber\\}

\newcommand{\ex}{\text{ex}}

\newcommand{\ed}{\end{document}}

\newcommand{\ie}{${i.e.,\ }$}

\begin{document}



\title{Chemical potential of a test hard sphere of variable size in a
hard-sphere fluid}


\author{David M. Heyes}
\email{david.heyes@rhul.ac.uk}
\affiliation{Department of Physics, University of London, Egham, Surrey TW20 0EX, UK}
\author{Andr\'es Santos}
\email{andres@unex.es}
\homepage{http://www.unex.es/eweb/fisteor/andres/}
\affiliation{Departamento de F\'{\i}sica and Instituto de Computaci\'on Cient\'ifica Avanzada (ICCAEx), Universidad de Extremadura,
E-06071 Badajoz, Spain}

\date{\today}

\begin{abstract}
The Lab\'ik and Smith Monte Carlo simulation technique to implement the Widom particle insertion method
is applied using Molecular Dynamics (MD) instead to calculate numerically the insertion probability, $P_0(\eta,\sigma_0)$, of tracer hard-sphere (HS) particles of different diameters, $\sigma_0$, in a host HS fluid of diameter $\sigma$ and packing fraction, $\eta$, up to $0.5$.
It is shown analytically that the only polynomial representation of $-\ln P_0(\eta,\sigma_0)$ consistent with the limits $\sigma_0\to 0$ and $\sigma_0\to\infty$ has necessarily a cubic form,
$c_0(\eta)+c_1(\eta)\sigma_0/\sigma+c_2(\eta)(\sigma_0/\sigma)^2+c_3(\eta)(\sigma_0/\sigma)^3$. Our MD data for $-\ln P_0(\eta,\sigma_0)$ are fitted to such a cubic polynomial and the functions
$c_0(\eta)$ and $c_1(\eta)$ are found to be statistically indistinguishable from their exact solution forms. Similarly, $c_2(\eta)$ and $c_3(\eta)$ agree very well with
the Boubl\'ik--Mansoori--Carnahan--Starling--Leland and Boubl\'ik--Carnahan--Starling--Kolafa formulas. The cubic polynomial is extrapolated (high density) or interpolated (low density) to obtain the chemical potential of the host fluid, or $\sigma_{0}\to\sigma$, as $\beta\mu^{\text{ex}}=c_0+c_1+c_2+c_3$. Excellent agreement between the Carnahan--Starling and Carnahan--Starling--Kolafa theories with our MD data is evident.

\end{abstract}

\date{\today}

\maketitle


\section{Introduction}
\label{sec1}

The statistical mechanical theory of hard-sphere (HS) fluids and solids is
important  as it underpins the phase behavior
and physical properties of a wide range of condensed phase
systems such as simple liquids, glasses, colloidal particles, emulsion droplets,
and granular materials.\cite{M08}
This work {reports Molecular Dynamics (MD)} simulations to test accurate analytic expressions
for the chemical potential of a HS impurity of variable diameter
at infinite dilution in a HS fluid. This information
is a useful precursor for understanding tracer solubility and HS mixtures in general. \\

We consider a test (or impurity) HS of diameter $\sigma_0$ immersed in a
sea of HSs of diameter $\sigma$ at a packing fraction $\eta$.\cite{BT16}
The quantity of interest here is the excess chemical potential of the test
particle, $\mu_0^\ex(\eta,\sigma_0)$, which becomes identical to the excess chemical potential $\mu^\ex(\eta)$ of the host fluid in the limit $\sigma_0\to\sigma$, \ie $\lim_{\sigma_0\to\sigma}\mu^\ex_0(\eta,\sigma_0)=\mu^\ex(\eta)$. As proved by Widom,\cite{W63b,SG81,DSD89}
the probability $P_0(\eta,\sigma_0)$ of successful insertion of the
test particle is related to the chemical potential {through}
\beq
P_0(\eta,\sigma_0)=e^{-\beta\mu_0^\ex(\eta,\sigma_0)},
\label{1.1}
\eeq
where $\beta=1/k_{B}T$ and $k_{B}$ is Boltzmann's constant.

The particle insertion technique has been applied to HS fluids for many
decades.\cite{HP62,W63b,NK91,A93,LS94,LJMS95,BMLS96,BT16}
However, if $\eta$ is rather large and $\sigma_0=\sigma$, the insertion probability is so small that the method becomes inefficient to measure directly $\mu^\ex(\eta)$ in computer simulations. In those situations, a circumventing path is needed.

Lab\'{i}k and Smith (LS)\cite{LS94} {proposed} a NVT Monte Carlo (MC) simulation technique
which can achieve this $\sigma_0\to\sigma$ limit accurately even at high densities.
The method measures the probability of the successful insertion of
a solute particle with a range of diameter values, $\sigma_{0}$, smaller than that of the solvent
HS diameter.  These measurements are extrapolated with a suitable polynomial
in powers of $\sigma_{0}$ to
$\sigma_0\to\sigma$, giving the chemical potential of the HS solvent.
{\it Inter alia} they give the tracer chemical potential of the test HS particle
 of diameter $\sigma_0 <\sigma$. The technique was subsequently extended to fused HS diatomics\cite{LJMS95} and
HS mixtures.\cite{BMLS96}

We note that recently
Baranau and Tallarek (BT)\cite{BT16} applied a solution consisting of measuring the so-called pore-size distribution, fitting it to
a Gaussian, and then performing analytically the integral in their Eq.~(11)
to finally determine the chemical potential.
This
is an alternative route to the chemical potential of the test
particle in the $\sigma_0\to\sigma$ limit.

In this work we follow instead the LS method
to calculate numerically
the insertion probability, $P_0(\eta,\sigma_0)$, for
different tracer HS sizes $\sigma_0$, {in a host HS fluid simulated by MD}.
The simulation obtained $-\ln P_0(\eta,\sigma_0)$ values are fitted to a cubic polynomial
$c_0(\eta)+c_1(\eta)\sigma_0/\sigma+c_2(\eta)(\sigma_0/\sigma)^2+c_3(\eta)(\sigma_0/\sigma)^3$
(a test function supported by several approximations), and then this polynomial is used
to extrapolate (high density) or interpolate (low density) to the value of this quantity at the desired diameter $\sigma$.
As mentioned above, a bonus from this way is that we obtain the chemical
potential $\mu_0(\eta,\sigma_0)$ for a tracer particle {with a diameter} both smaller
and (for some densities) also larger than $\sigma$ (not only for a fluid particle of the
same size as the host fluid HSs).
The density-dependent coefficients $c_n$ are also determined, which enables a more
detailed comparison with theoretical predictions to
be made. Instead of comparing
only the chemical potential of the host fluid particle (\ie $c_{0}+c_{1}+c_{2}+c_{3}$) as a
function of density (as was done, for instance, in Fig.~1(a) of BT's paper), we
validate the accuracy of the simulations by (i) confirming agreement with the exact $c_0$
and $c_1$ and (ii) comparing two extra coefficients ($c_2$ and $c_3$) with literature
theoretical predictions, which builds on the pioneering {LS} work.\cite{LS94} \\

{The remainder of this paper is organized as follows. The standard theoretical approximations are reviewed in Sec.~\ref{sec2} and the use of a cubic polynomial as a trial function for $\beta\mu_0^\ex$ is justified. Section \ref{sec3} summarizes the Widom particle insertion method and describes the way it is implemented in our MD simulations. The results are presented and compared with theoretical predictions in Sec.~\ref{sec4}. Finally, the paper is closed with some conclusions in Sec.~\ref{sec5}.}


\section{Theoretical approximations}
\label{sec2}
\subsection{Multi-component hard-sphere fluids}

Let us start by considering a (three-dimensional) fluid mixture of additive HSs with an arbitrary number of components. There are $N_j$  spheres of species $j$  having a diameter $\sigma_j$, so that the total number of particles is $N=\sum_j N_j$ and the $n$th moment of the size distribution is
\beq
M_n=\frac{\sum_{j}N_j\sigma_j^n}{N}.
\label{2.2}
\eeq
The total packing fraction is
\beq
\eta=\frac{\pi}{6}\frac{N}{V}M_3,
\label{eta}
\eeq
where $V$ is the volume of the system.

We will denote the compressibility factor of the mixture by $Z(\eta,\{N_j\})\equiv p V/N k_BT$, where $p$ is the pressure. Since its exact form is not known,  several approximations have been proposed.\cite{MGPC08,BS08} In particular, the exact solution\cite{LZ71,PS75,B75} of the Percus--Yevick (PY) integral equation\cite{PY58} allows one to obtain explicit expressions for $Z(\eta,\{N_j\})$ through different thermodynamic routes.
The virial (PY-v), compressibility (PY-c), and chemical-potential (PY-$\mu$) routes in the PY approximation share the following common structure:\cite{S16,LZ71,PS75,B75,S12b,SR13}
\beq
Z(\eta,\{N_j\})=Z_0(\eta)+Z_1(\eta)\frac{M_1M_2}{M_3}+Z_2(\eta)\frac{M_2^3}{M_3^2},
\label{2.1}
\eeq
where
\beq
Z_0(\eta)=\frac{1}{1-\eta},\quad
Z_1(\eta)=\frac{3\eta}{(1-\eta)^2}.
\label{2.3}
\eeq
The coefficient $Z_2(\eta)$ depends on the route and several literature predictions are
displayed in Table \ref{table1}. On the other hand, the coefficients \eqref{2.3} are the same in all the PY approximations. As will be discussed later (see {also} the Appendix), those coefficients are exact.

Since none of the three prescriptions (PY-v, PY-c, and PY-$\mu$) is particularly accurate, Boubl\'ik\cite{B70} and, independently, Mansoori \emph{et al}.\cite{MCSL71} proposed an interpolation between PY-v and PY-c with respective weights $1/3$ and $2/3$. The resulting Boubli\'ik--Mansoori--Carnahan--Starling--Leland (BMCSL) compressibility factor has of course the structure \eqref{2.1} with $Z_0$ and $Z_1$ given by Eq.~\eqref{2.3}
and the corresponding expression for $Z_2$ is also included in Table \ref{table1}.
In the monodisperse case (\ie $\sigma_j\to\sigma\Rightarrow M_n\to \sigma^n$) {one has} $Z=Z_0+Z_1+Z_2$, and the BMCSL equation of state reduces to the Carnahan--Starling (CS) one,\cite{CS69,S16,HCPE07}
\begin{equation}
Z_{\text{CS}}(\eta)=\frac{1+\eta+\eta^{2}-\eta^{3}}{(1-\eta)^{3} }.
\label{eq:xxaas1}
\end{equation}
In 1986, Kolafa proposed a slight correction to the CS equation, namely
\begin{equation}
Z_{\text{CSK}}(\eta)=\frac{1+\eta+\eta^{2}-\frac{2}{3}\eta^{3}(1+\eta)}{(1-\eta)^{3} }.
\label{eq:xxaas2}
\end{equation}
It first appeared as Eq.~(4.46) in a review paper by Boubl{\'\i }k and  Nezbeda.\cite{BN86} Following Kolafa's recommendation,\cite{K98} we will refer to Eq.~\eqref{eq:xxaas2} as the Carnahan--Starling--Kolafa (CSK) equation of state. The extension of $Z_{\text{CSK}}$ to mixtures was carried out by Boubl{\'\i }k\cite{B86} by keeping the structure \eqref{2.1} and choosing $Z_2$ as $Z_2=Z_{\text{CSK}}-Z_0-Z_1$. The resulting Boubl\'ik--Carnahan--Starling--Kolafa (BCSK) expression is given in the bottom row of Table \ref{table1}.

The excess free energy per particle of the mixture, $a^\ex(\eta,\{N_j\})$, is related to the compressibility factor $Z(\eta,\{N_j\})$ through\cite{S16}
\beq
\beta a^\ex(\eta,\{N_j\})= \int_0^1 dt\frac{Z(\eta t,\{N_j\})-1}{t}.
\label{2.4}
\eeq
Therefore, the class of approximations of the form \eqref{2.1} yield
\beq
\beta a^\ex(\eta,\{N_j\})=c_0(\eta)+c_1(\eta)\frac{M_1M_2}{M_3}+a_2(\eta)\frac{M_2^3}{M_3^2},
\label{2.5}
\eeq
where{
\begin{subequations}
\label{2.6&2.4b}
\beq
Z_0(\eta)=1+\eta c_0'(\eta)\Rightarrow c_0(\eta)=-\ln(1-\eta),
\label{2.6.0}
\eeq
\beq
Z_1(\eta)=\eta c_1'(\eta)\Rightarrow  c_1(\eta)=\frac{3\eta}{1-\eta},
\label{2.6.1}
\eeq
\beq
Z_2(\eta)=\eta a_2'(\eta)\Rightarrow a_2(\eta)= \int_0^1 dt\frac{Z_2(\eta t)}{t},
\label{2.4b}
\eeq
\end{subequations}
the primes denoting derivatives with respect to $\eta$.}
The expressions for the coefficient $a_2(\eta)$ corresponding to the approximations PY-v, PY-c, PY-$\mu$, BMCSL, and BCSK are also included in Table \ref{table1}.

\begin{table}
   \caption{Expressions of $Z_2(\eta)$ [see Eq.~\eqref{2.1}] and $a_2(\eta)$ [see Eq.~\eqref{2.5}] according to several approximations.}\label{table1}
\begin{ruledtabular}
\begin{tabular}{ccc}
Approx.&$Z_2(\eta)$&$a_2(\eta)$\\
\hline
PY-v&$\displaystyle{\frac{3\eta^2}{(1-\eta)^2}}$&
$\displaystyle{3\ln(1-\eta)+\frac{3\eta}{1-\eta}}$\\
PY-c&$\displaystyle{\frac{3\eta^2}{(1-\eta)^3}}$&
$\displaystyle{\frac{3\eta^2}{2(1-\eta)^2}}$\\
PY-$\mu$&$\displaystyle{-\frac{9\ln(1-\eta)}{\eta}-9\frac{1-\frac{3}{2}\eta}{(1-\eta)^2}}$&
$\displaystyle{\frac{9\ln(1-\eta)}{\eta}+9\frac{1-\frac{1}{2}\eta}{1-\eta}}$\\
BMCSL&$\displaystyle{\frac{\eta^2(3-\eta)}{(1-\eta)^3}}$&$\displaystyle{\ln(1-\eta)+\frac{\eta}{(1-\eta)^2}}$\\
BCSK&$\displaystyle{\frac{\eta^2[3-\frac{2}{3}\eta(1+\eta)]}{(1-\eta)^3}}$&
$\displaystyle{\frac{8}{3}\ln(1-\eta)}$\\
 & & $\displaystyle{+\eta\frac{16-15\eta+4\eta^2}{6(1-\eta)^2}}$\\
   \end{tabular}
 \end{ruledtabular}
 \end{table}

We now consider the excess chemical potential of a generic species $i$,
which is thermodynamically defined as\cite{S16}
\beq
\mu_i^\ex=\left(\frac{\partial N a^\ex}{\partial N_i}\right)_{V,N_{j\neq i}}.
\label{2.7}
\eeq
In order to take the derivative in Eq.~\eqref{2.5}, we need to make use of the mathematical properties
\begin{subequations}
\beq
N\left(\frac{\partial \eta}{\partial N_i}\right)_{V,N_{j\neq i}}=\eta\frac{\sigma_i^3}{M_3},
\label{2.8}
\eeq
\bal
N\left(\frac{\partial N M_1 M_2/M_3}{\partial N_i}\right)_{V,N_{j\neq i}}=&\frac{M_1M_2}{M_3}\left(\frac{\sigma_i}{M_1}+\frac{\sigma_i^2}{M_2}\right.\nn
&\left.-\frac{\sigma_i^3}{M_3}\right),
\label{2.9}
\eal
\beq
N\left(\frac{\partial N M_2^3/M_3^2}{\partial N_i}\right)_{V,N_{j\neq i}}=\frac{M_2^3}{M_3^2}\left(3\frac{\sigma_i^2}{M_2}-2\frac{\sigma_i^3}{M_3}\right).
\label{2.10}
\eeq
\end{subequations}
Therefore, the final result stemming from Eq.~\eqref{2.5} is
\bal
\beta \mu_i^\ex(\eta,\{N_j\})=&c_0(\eta)+c_1(\eta)\frac{M_1M_2}{M_3}\frac{\sigma_i}{M_1}\nn
&+
\left[c_1(\eta)\frac{M_1M_2}{M_3}+3a_2(\eta)\frac{M_2^3}{M_3^2}\right]\frac{\sigma_i^2}{M_2}\nn
&+\left\{\eta c_0'(\eta)+\left[\eta c_1'(\eta)-c_1(\eta)\right]\frac{M_1M_2}{M_3}\right.\nn
&\left.
+\left[\eta a_2'(\eta)-2a_2(\eta)\right]\frac{M_2^3}{M_3^2}\right\}\frac{\sigma_i^3}{M_3}.
\label{2.11}
\eal
Note that Eqs.~\eqref{2.1}, \eqref{2.5}, and \eqref{2.11} are consistent with the exact thermodynamic relation
\beq
\frac{1}{N}\sum_{i} N_i\beta\mu_i^\ex=\beta a^\ex+Z-1,
\label{2.11b}
\eeq
thanks to the {properties in \eqref{2.6&2.4b}}, regardless of the expression {for} $a_2(\eta)$.

{As proved in the Appendix (where a general dimensionality $d$ is considered),} Eq.~\eqref{2.11} is \emph{exact} to first order in $\sigma_i$, \ie
\beq
\beta \mu_i^\ex(\eta,\{N_j\})=c_0(\eta)+c_1(\eta)\frac{M_1M_2}{M_3}\frac{\sigma_i}{M_1}+\mathcal{O}(\sigma_i^2).
\label{2.11c}
\eeq
This in turn proves the exact character of the coefficients $c_0$ and $c_1$ in {Eqs.~\eqref{2.6.0} and \eqref{2.6.1}, respectively,} and, hence, of the coefficients $Z_0$ and $Z_1$ in Eq.~\eqref{2.3}, as anticipated before.

\subsection{Test particle in a one-component hard-sphere fluid}
In this special case, we can set $M_n\to \sigma^n$ and particularize Eq.~\eqref{2.11} to a species $i=0$ made of a single particle of diameter $\sigma_0$. The result is
\beq
\beta\mu_0^\ex(\eta,\sigma_0)=c_0(\eta)+c_1(\eta)\frac{\sigma_0}{\sigma}+c_2(\eta)\frac{\sigma_0^2}{\sigma^2}+c_3(\eta)\frac{\sigma_0^3}{\sigma^3},
\label{2.12}
\eeq
where
\begin{subequations}
\label{2.13&2.14}
\beq
c_2(\eta)=c_1(\eta)+3a_2(\eta),
\label{2.13}
\eeq
\beq
c_3(\eta)=\eta c_0'(\eta)+\eta c_1'(\eta)-c_1(\eta)
+\eta a_2'(\eta)-2a_2(\eta).
\label{2.14}
\eeq
\end{subequations}
{Notice that from Eqs.~\eqref{2.6&2.4b} and \eqref{2.13&2.14} one can obtain the simple relation\cite{LS94}
\beq
c_3(\eta)=Z(\eta)-1-\frac{1}{3}c_1(\eta)-\frac{2}{3}c_2(\eta).
\label{LSeq}
\eeq
}

Inserting {Eqs.~\eqref{2.6.0} and \eqref{2.6.1} together with} the approximate expressions of $a_2$ listed in Table \ref{table1} into Eqs.~\eqref{2.13&2.14}, one can obtain the approximate expressions for the coefficients $c_2$ and $c_3$ given in Table \ref{table2}. The last column of Table \ref{table2} {presents formulas for} the excess chemical potential of the fluid, \ie $\beta\mu^{\ex}(\eta)=\lim_{\sigma_0\to 1}\beta\mu_0^\ex(\eta,\sigma_0)=c_0(\eta)+c_1(\eta)+c_2(\eta)+c_3(\eta)$,
{for the various approximations}.

\begin{table*}
   \caption{Expressions of $c_2(\eta)$, $c_3(\eta)$ [see Eq.~\eqref{2.12}], and $\beta\mu^{\ex}(\eta)$ according to several approximations.}\label{table2}
\begin{ruledtabular}
\begin{tabular}{cccc}
Approx.&$c_2(\eta)$&$c_3(\eta)$&$\beta\mu^{\ex}(\eta)$\\
\hline
PY-v&$\displaystyle{9\ln(1-\eta)+12\frac{\eta}{1-\eta}}$&
$\displaystyle{-6\ln(1-\eta)-\eta\frac{5-11\eta}{(1-\eta)^2}}$&
$\displaystyle{2\ln(1-\eta)+2\eta\frac{5-2\eta}{(1-\eta)^2}}$\\
PY-c&$\displaystyle{3\eta\frac{2+\eta}{2(1-\eta)^2}}$&
$\displaystyle{\eta\frac{1+\eta+\eta^2}{(1-\eta)^3}}$&
$\displaystyle{-\ln(1-\eta)+\eta\frac{14-13\eta+5\eta^2}{2(1-\eta)^3}}$\\
PY-$\mu$&$\displaystyle{27\frac{\ln(1-\eta)}{\eta}+3\frac{18-7\eta}{2(1-\eta)}}$&
$\displaystyle{-27\frac{\ln(1-\eta)}{\eta}-\frac{54-83\eta+14\eta^2}{2(1-\eta)^2}}$&
$\displaystyle{-\ln(1-\eta)+\eta\frac{14+\eta}{2(1-\eta)^2}}$\\
BMCSL&$\displaystyle{3\ln(1-\eta)+3\eta\frac{2-\eta}{(1-\eta)^2}}$&$\displaystyle{-2\ln(1-\eta)-\eta\frac{1-6\eta+3\eta^2}{(1-\eta)^3}}$&$\displaystyle{\eta\frac{8-9\eta+3\eta^2}{(1-\eta)^3}}$\\
BCSK&$\displaystyle{8\ln(1-\eta)+\eta\frac{22-21\eta+4\eta^2}{2(1-\eta)^2}}$&
$\displaystyle{-\frac{16}{3}\ln(1-\eta)-\eta\frac{13-43\eta+27\eta^2-2\eta^3}{3(1-\eta)^3}}$&
$\displaystyle{\frac{5}{3}\ln(1-\eta)+\eta\frac{58-79\eta+39\eta^2-8\eta^3}{6(1-\eta)^3}}$\\
\end{tabular}
 \end{ruledtabular}
 \end{table*}

Given that a number of approximations (PY-v, PY-c, PY-$\mu$, BMCSL, and BCSK) share the common cubic polynomial form \eqref{2.12} (with the exact coefficients $c_0$ and $c_1$) for the excess chemical potential of a test particle immersed in a monodisperse HS fluid, one might reasonably
query whether {one could construct either} a simpler approximation (with adjustable $c_2$) {from} a quadratic polynomial or
a more accurate approximation (with adjustable $c_2$, $c_3$, $c_4$, \ldots) from  a polynomial of degree higher than three.
However, as we
will
see, if $\beta \mu_0^\ex(\eta,\sigma_0)$ is represented by a polynomial in the diameter $\sigma_0$, the polynomial must \emph{necessarily} be of third degree.
This is a consequence of the physical requirement that, in the limit of an infinitely large impurity, one must have\cite{RFHL60,RELK02,S12c}
\beq
\eta Z(\eta)=\lim_{\sigma_0\to\infty}\frac{\beta\mu_0^\ex(\eta,\sigma_0)}{(\sigma_0/\sigma)^3}.
\label{2b.4}
\eeq
Therefore, since $\lim_{\sigma_0\to\infty}{\beta\mu_0^\ex(\eta,\sigma_0)}/{\sigma_0^3}$ can be neither zero nor infinity, the only polynomial approximations consistent with that property are third-degree ones.

In the case of the approximations of the form \eqref{2.1}, Eq.~\eqref{2b.4} implies
\beq
c_3(\eta)=\eta Z(\eta)=\eta\left[Z_0(\eta)+Z_1(\eta)+Z_2(\eta)\right].
\label{2c.5}
\eeq
{It can be noticed} that Eq.~\eqref{2c.5} is independent of {Eq.~\eqref{LSeq}}. In fact,
it can be easily checked that the PY-v, PY-$\mu$, BMCSL, and BCSK expressions for $Z_2(\eta)$ (see Table \ref{table1}) and $c_3(\eta)$ (see Table \ref{table2}) are inconsistent with Eq.~\eqref{2c.5}.
This means that those approximations \emph{qualitatively} agree with the physical requirement \eqref{2b.4} in that $\lim_{\sigma_0\to\infty}{\beta\mu_0^\ex(\eta,\sigma_0)}/{\sigma_0^3}=\text{finite}$ but yield different results for the left- and right-hand sides. On the other hand, the PY-c approximation, which actually is equivalent to the Scaled Particle Theory (SPT) approximation,\cite{RFL59,LHP65,MR75,R88,HC04} is fully consistent with Eqs.~\eqref{2b.4} and \eqref{2c.5}. As a matter of fact, the PY-c/SPT cubic prescription for $\beta\mu_{0}^\ex(\eta,\sigma_0)$ is the only one that is simultaneously consistent with both Eqs.~\eqref{LSeq} and \eqref{2c.5} without violating the value $b_3=10$ for the  third virial coefficient of the one-component fluid. Combination of Eqs.~\eqref{LSeq} and \eqref{2c.5} {[together with Eqs.~\eqref{2.6&2.4b} and \eqref{2.13}]} yields the differential equation $a_2'(\eta)=2a_2(\eta)/\eta(1-\eta)$, whose general solution is $a_2(\eta)=K\eta^2/(1-\eta)^2$, $K$ being a constant. The associated third virial coefficient is $b_3=7+2K$, so that $b_3=10$ implies $K=\frac{3}{2}$ and thus one recovers the PY-c/SPT approximation.

Section \ref{sec3} describes the process and results of a MD simulation
study of this HS system which were carried out to help determine
which of the approximations for $c_{2}$ and $c_{3}$ (see Table \ref{table2}) is best.\\


\section{Widom's particle insertion method and Molecular Dynamics simulation}
\label{sec3}

Consider an $N$-particle system where $\Phi_N(\mathbf{r}^{N})$ is
the potential energy. The Widom
particle insertion method for the excess chemical potential
$\mu^{\ex}$  is\cite{HP62,W63b,HCD90,H92}
\bal
e^{-\beta\mu^{\ex}}=&\frac{\int d\mathbf{r}^{N+1} \,e^{-\beta\Phi_N(\mathbf{r}^N)}e^{-\beta\Delta\Phi_{N+1}(\mathbf{r}^{N+1})}}{V\int d\mathbf{r}^N \,e^{-\beta\Phi_N(\mathbf{r}^N)}}\nn
=&
\left\langle e^{-\beta\Delta\Phi_{N+1}(\mathbf{r}^{N+1})}\right\rangle,
\label{eq:nkzaa2a}
\eal
where
$\Delta \Phi_{N+1}(\mathbf{r}^{N+1})= \Phi_{N+1}(\mathbf{r}^{N+1})-\Phi_N(\mathbf{r}^{N})$ and the ensemble average is denoted by $\langle\cdots\rangle$.
The $(N+1)$th particle (here denoted by the subscript $0$) can be considered to be a test particle, as
it does not influence the physical distribution of the
other $N$ particles. Hence,
\begin{equation}
\beta\mu_0^{\ex}=-\ln\left\langle e^{-\beta\Delta\Phi_{N+1}(\mathbf{r}^{N+1})}\right\rangle.
\label{eq:xxaas}
\end{equation}

The test particle is inserted randomly into the $N$-particle host fluid.
The important point is that it does so in a non-intrusive way.
For HSs, Eq.~(\ref{eq:xxaas}) reduces
to a simple bookkeeping procedure as $\exp(-\beta\Delta\Phi_{N+1})$ {either is} $1$ when the
test sphere does not overlap with any of the $N$ particles or is equal to $0$
if it overlaps with any of them. As discussed in Sec.\ \ref{sec2}, the test particle does not need to be the same type of particle as the
other $N$ particles. We consider particle $\alpha=0$ to be an impurity
HS of diameter $\sigma_{0}$, taking the HS diameter of the host fluid
to be $\sigma$.

Our numerical implementation of the Widom insertion method run as follows. At a given packing fraction $\eta$, a monodisperse HS fluid  was simulated by a standard MD method. The procedure was to randomly
insert a test ``point'' in the system and calculate the distance $r_{n}$ from that point to the center of the nearest sphere. Then, all the values from $\sigma_0=0$ to $\sigma_0=2r_{n}-\sigma$ represented  accepted insertions, which were
accumulated efficiently in a histogram at the same time in the MD
simulation. In addition, as  the test particles
are introduced in a non-intrusive way, many of them can be inserted at the same time, and we
used the same number of test particles as the number of host
fluid particles. One difference with the LS method\cite{LS94} is {that} we use MD
rather than MC to evolve the host fluid assembly of HSs.\\

\begin{figure}[tbp]
\includegraphics[width=8cm]{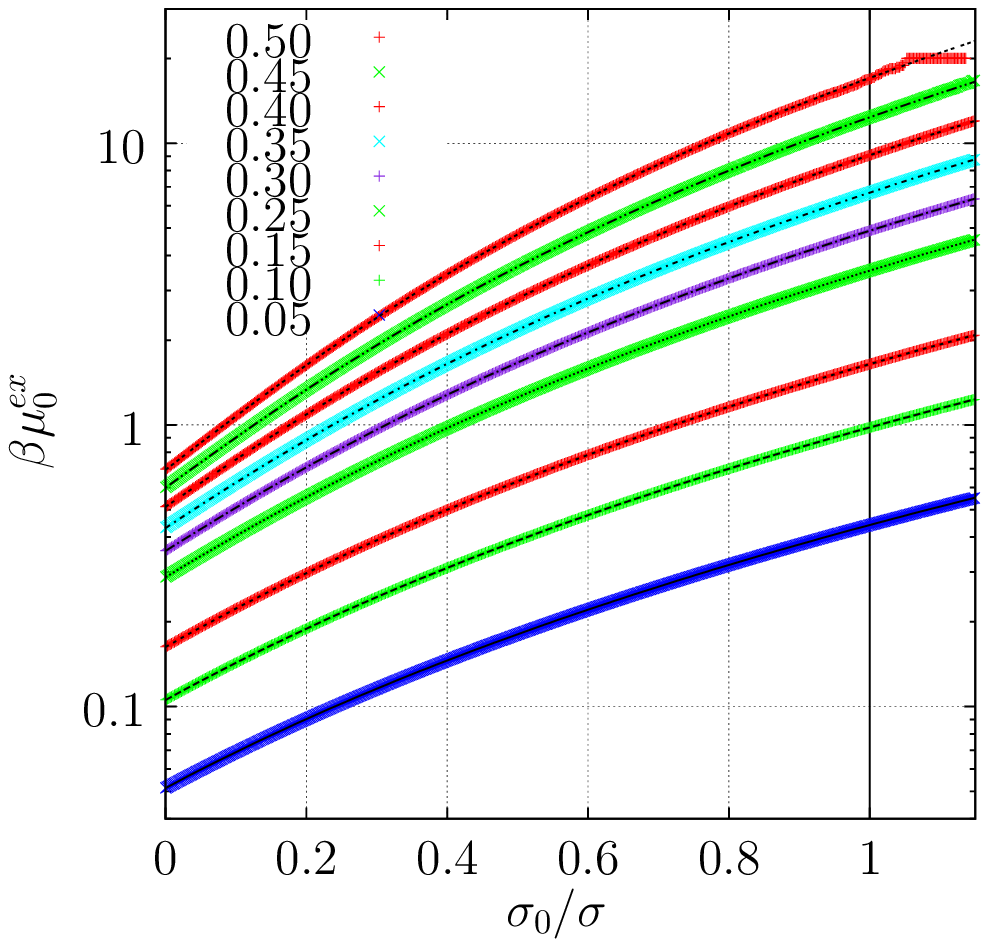}
\caption{Plot of the excess chemical potential of a test particle, $\beta\mu_0^\ex(\eta,\sigma_0)$, as a function of the diameter $\sigma_0$. The symbols are the values obtained in our MD simulations by the Widom insertion method, while the lines are least-square fits to cubic polynomials of the form \eqref{2.12} with free coefficients $c_n$.
The noisiest data for large $\eta$ and $\sigma_0$ were excluded from the fits. The different values of $\eta$ are indicated in the legend. }
\label{fig1}
\end{figure}

For each trial insertion $r_{n}$,   $1$ was added to all entrants of a
histogram (rather like that for the radial
distribution function) {for} $P_0(\eta,\sigma_0)$ for $\sigma_0=2r_{n}-\sigma$ and all $\sigma_0$ values less than $2r_{n}-\sigma$
at the same time. This is a statistically efficient procedure for computing the chemical potential of
the impurity at infinite dilution,
$\beta\mu^{\ex}_0(\eta,\sigma_{0})$. The chemical potential of the
HS fluid is just  $\mu^{\ex}(\eta)=\mu^\ex_0(\eta,\sigma)$
when $\sigma_{0}=\sigma$. At not too high densities, data on the
chemical potential for $\sigma_{0}>\sigma$ can also be obtained, and so the
HS chemical potential becomes a matter of interpolation and data fitting in that case.
For states near a packing fraction $\eta\approx 0.50$ the HS chemical potential
needs to be estimated by extrapolation of the $\sigma_{0}<\sigma$
histogram entrants, as the probability of inserting
a HS in a HS fluid during a typical simulation can be
impracticably small (less than $10^{-7}$).\\

At each density, the MD values of $\beta\mu_0(\eta,\sigma_0)$ as a function of $\sigma_0$ were fitted to the cubic polynomial \eqref{2.12} to obtain the four coefficients $c_0$--$c_3$,  without imposing the exact values \eqref{2.6.0} and \eqref{2.6.1} of $c_0$ and $c_1$.
This contrasts with the LS procedure,\cite{LS94} where the coefficients $c_0$ and $c_1$ were fixed to be given by Eqs.~\eqref{2.6.0} and \eqref{2.6.1}, the coefficient $c_3$ was forced to satisfy the relationship \eqref{LSeq} (with $Z$ obtained by independent MC simulations of the host fluid), and therefore only the coefficient $c_2$ was fitted to the simulation data of $-\ln P_0(\eta,\sigma_0)$. In addition, the maximum value of $\sigma_0$ used in the least-square fitting corresponded to\cite{LS94} $P_0\approx 10^{-3}$.

{Our} simulations were carried out with $N=2048$ HSs. There were {ca.} $1.4 \times 10^{5}$ collisions per particle at $\eta=0.05$ and
$5.6 \times 10^{5}$ collisions per particle at $\eta=0.5$.
The maximum value of $\sigma_0$ chosen for the fitting process was $1.10\sigma$,
for $\eta < 0.4$, decreasing to $0.90\sigma$ for $\eta=0.46$ to $0.80\sigma$ for $\eta\ge 0.48$. {This corresponded to $P_0\approx 2\times 10^{-5}$.}
The insertion probability histogram had a resolution of $0.005\sigma$.

\section{Results}
\label{sec4}

Figure \ref{fig1} shows the values of $\beta\mu_0^\ex(\eta,\sigma_0)$ obtained in our simulations for nine representative packing fractions from $\eta=0.05$ to $\eta=0.50$. The least-square fits to a cubic polynomial are also included in Fig.~\ref{fig1} and an excellent agreement is found.

\begin{figure}[tbp]
\includegraphics[width=8cm]{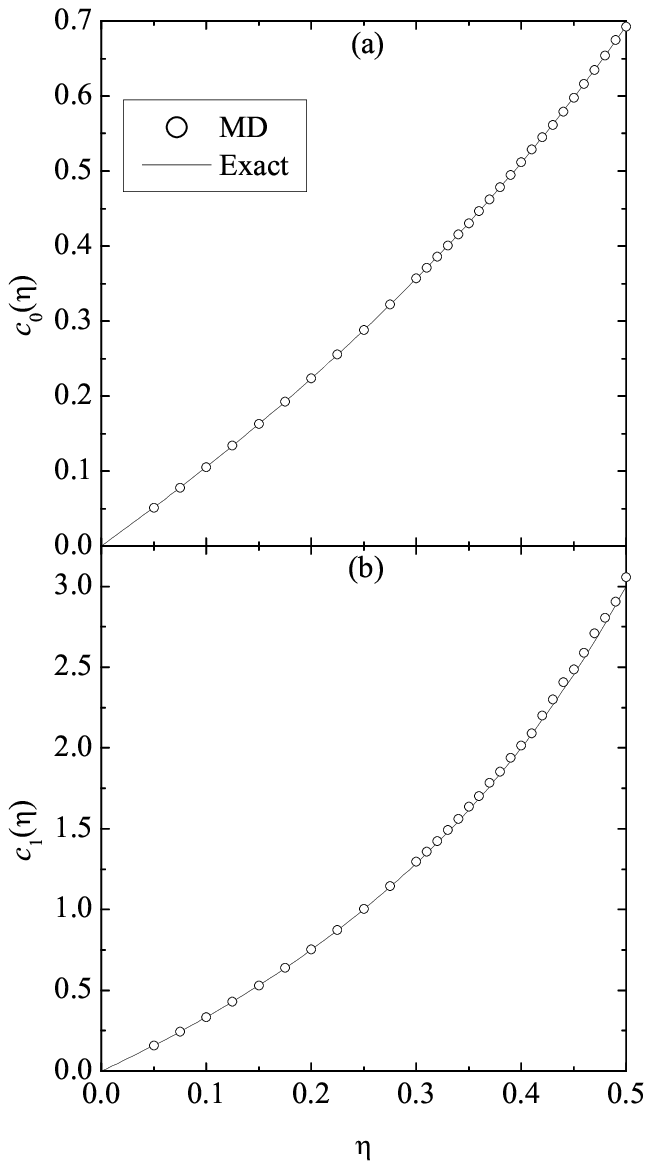}
\caption{Plot of the coefficients (a) $c_0(\eta)$ and (b) $c_1(\eta)$. The lines represent the exact expressions [see {Eqs.~\eqref{2.6.0} and \eqref{2.6.1}}], while the symbols represent the values obtained from a least-square fit of MD data.}
\label{c0_c1}
\end{figure}

The extracted values of the coefficients $c_0(\eta)$ and $c_1(\eta)$ are plotted in Fig.~\ref{c0_c1} for $31$ values of $\eta$ ranging from $0.05$ to $0.50$. Comparison with the exact expressions {\eqref{2.6.0} and \eqref{2.6.1}} shows an extremely good agreement. This confirms and reinforces the reliability and accuracy of our MD results.

\begin{figure}[tbp]
\includegraphics[width=8cm]{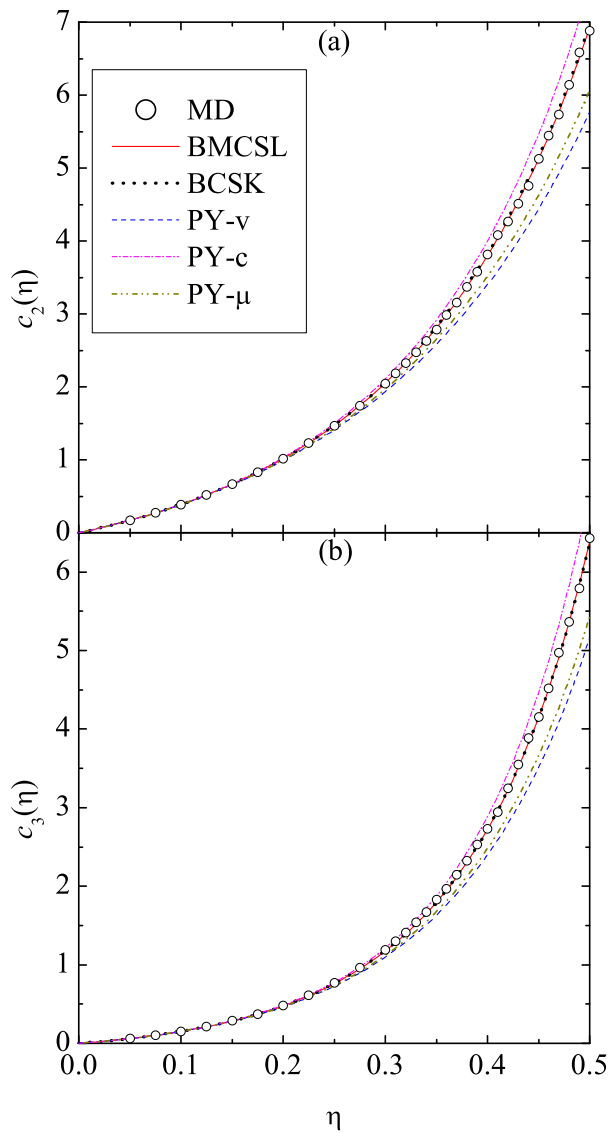}
\caption{Plot of the coefficients (a) $c_2(\eta)$ and (b) $c_3(\eta)$. The lines represent the theoretical expressions (see Table \ref{table2}), while the symbols represent the values obtained from a least-square fit of MD data.}
\label{c2_c3}
\end{figure}

Figure \ref{c2_c3} displays the values of the fitted coefficients $c_2(\eta)$ and $c_3(\eta)$ for the same densities as in Fig.~\ref{c0_c1}. Since the exact expressions of $c_2$ and $c_3$ are (to the best of our knowledge) unknown, we compare the simulation values with the approximate theoretical predictions considered in Table \ref{table2}. Up to $\eta\simeq 0.2$ all the theories practically overlap and reproduce the MD values. At higher densities, however, the three PY predictions clearly deviate from the simulation data: while the PY-c approximation overestimates the data, the PY-$\mu$ and, especially, the PY-v approximations underestimate them. On the other hand, the BMCSL and BCSK curves, which are practically indistinguishable, reproduce excellently the MD results.

\begin{figure}[tbp]
\includegraphics[width=8cm]{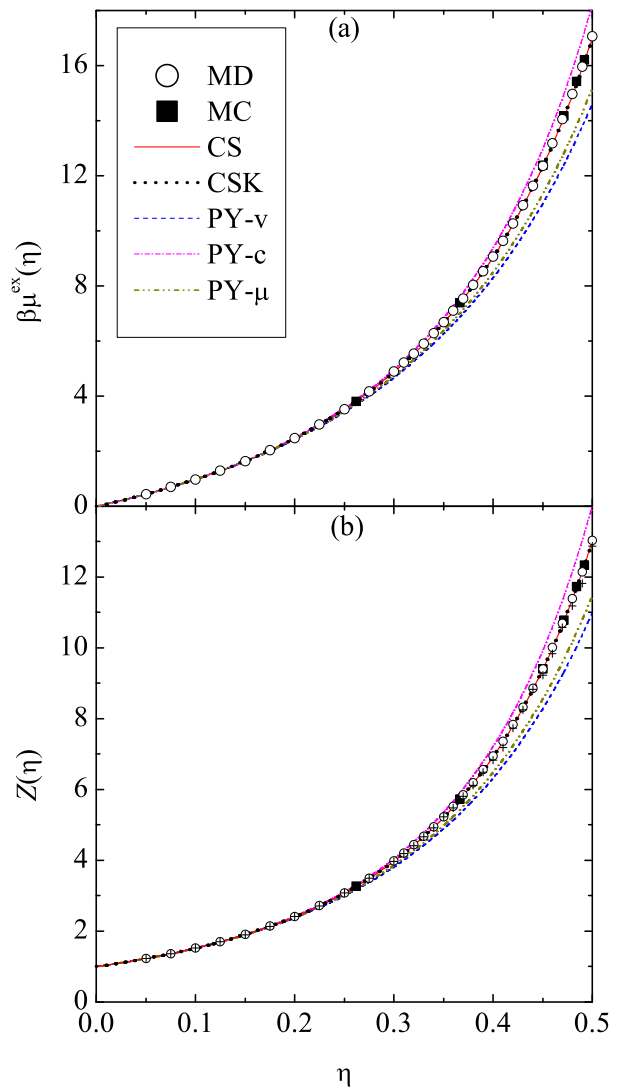}
\caption{{Plot of (a) the excess chemical potential $\beta\mu^\ex(\eta)$ and (b) the compressibility factor $Z(\eta)$. The lines represent the theoretical expressions (see Tables \ref{table1} and  \ref{table2}), the open circles represent $\beta\mu^\ex=c_0+c_1+c_2+c_3$ and $Z=1+\frac{1}{3}c_1+\frac{2}{3}c_2+c_3$ (with coefficients $c_n$ obtained from a least-square fit of our MD data), and the filled squares represent the MC data of Ref.~\onlinecite{LS94}. In panel (b), the crosses represent $c_3/\eta$.}}
\label{mu}
\end{figure}

\begingroup\squeezetable
\begin{table*}
   \caption{{Numerical values of $c_0$, $c_1$, $c_2$, $c_3$, $\beta\mu^{\ex}=c_0+c_1+c_2+c_3$,
and $Z=1+\frac{1}{3}c_1+\frac{2}{3}c_2+c_3$, as obtained from our MD simulations.
{The errors in the constants $c_0$--$c_3$ range from $0.002$--$0.01$\% at $\eta=0.075$
to $0.06$--$0.15$\% at $\eta=0.50$.}}}\label{table3}
\begin{ruledtabular}
\begin{tabular}{ddddddd}
\multicolumn{1}{c}{$\;\;\;\;\;\eta$}&\multicolumn{1}{c}{$\;\;\;\;\;c_0$}&\multicolumn{1}{c}{$\;\;\;\;\;c_1$}&\multicolumn{1}{c}{$\;\;\;\;\;c_2$}&\multicolumn{1}{c}{$\;\;\;\;\;c_3$}&
\multicolumn{1}{c}{$\;\;\;\;\;\beta\mu^{\ex}$}&\multicolumn{1}{c}{$\;\;\;\;\;Z$}\\
\hline
0.050&0.05138&0.15791&0.17053&0.06132&0.4411&1.2276\\
0.075&0.07807&0.24363&0.27251&0.10232&0.6965&1.3652\\
0.100&0.10551&0.33413&0.38726&0.15254&0.9794&1.5221\\
0.125&0.13371&0.42976&0.51731&0.21300&1.2938&1.7011\\
0.150&0.16277&0.53066&0.66523&0.28545&1.6441&1.9058\\
0.175&0.19264&0.63853&0.83044&0.37440&2.0360&2.1409\\
0.200&0.22340&0.75338&1.01629&0.48254&2.4756&2.4112\\
0.225&0.25519&0.87520&1.22927&0.61128&2.9709&2.7225\\
0.250&0.28802&1.00569&1.46854&0.76883&3.5311&3.0831\\
0.275&0.32190&1.14594&1.73817&0.96082&4.1668&3.5016\\
0.300&0.35694&1.29667&2.04555&1.19320&4.8924&3.9891\\
0.310&0.37134&1.35932&2.18212&1.29733&5.2101&4.2052\\
0.320&0.38594&1.42389&2.32556&1.41204&5.5474&4.4370\\
0.330&0.40065&1.49284&2.47122&1.53868&5.9034&4.6838\\
0.340&0.41582&1.56042&2.63206&1.67237&6.2807&4.9472\\
0.350&0.43065&1.63751&2.78336&1.82938&6.6809&5.2308\\
0.360&0.44664&1.70377&2.98153&1.97059&7.1025&5.5262\\
0.370&0.46210&1.78329&3.15755&2.14789&7.5508&5.8473\\
0.380&0.47863&1.85519&3.37150&2.32390&8.0292&6.1900\\
0.390&0.49464&1.93913&3.57437&2.52817&8.5363&6.5575\\
0.400&0.51157&2.01626&3.81678&2.72871&9.0733&6.9453\\
0.410&0.52923&2.09101&4.08354&2.94310&9.6469&7.3625\\
0.420&0.54483&2.20251&4.26902&3.24896&10.265&7.8291\\
0.430&0.56175&2.30181&4.51157&3.54899&10.924&8.3240\\
0.440&0.57897&2.40785&4.75777&3.88409&11.629&8.8586\\
0.450&0.59808&2.48743&5.12705&4.15586&12.368&9.4030\\
0.460&0.61655&2.58968&5.44313&4.52232&13.172&10.014\\
0.470&0.63462&2.71022&5.73096&4.97255&14.048&10.697\\
0.480&0.65443&2.80728&6.14140&5.36583&14.969&11.396\\
0.490&0.67473&2.90778&6.58217&5.79039&15.955&12.148\\
0.500&0.69285&3.05778&6.87840&6.43386&17.063&13.039\\
\end{tabular}
\end{ruledtabular}
\end{table*}
\endgroup

Now that we have validated our numerical values of the four coefficients $c_n$ characterizing the diameter dependence of the impurity chemical potential $\beta\mu_0^\ex$,
an accurate estimate of the chemical potential of the pure HS fluid, written as
$\beta\mu^\ex=c_0+c_1+c_2+c_3$, can be made. The results are shown in Fig.~\ref{mu}(a), where they are compared with the PY, CS, and CSK approximations (see again Table \ref{table2}).
The observed trends are similar to those presented in Fig.~\ref{c2_c3}. In particular,
there is excellent agreement between the present MD results and the CS and CSK theories. {Figure \ref{mu}(a) also includes the MC data reported in Ref.~\onlinecite{LS94}, which are fully consistent with our MD results.}

{An interesting additional feature of our approach is that we can predict the compressibility factor $Z(\eta)$ of the HS fluid via Eq.~\eqref{LSeq} from the knowledge of the coefficients $c_n$ characterizing the size dependence of the solute chemical potential $\beta\mu^\ex_0$, \ie $Z=1+\frac{1}{3}c_1+\frac{2}{3}c_2+c_3$. This quantity is plotted in Fig.~\ref{mu}(b), where it shows again an excellent agreement with the CS and CSK approximations, as well as with the results obtained in Ref.~\onlinecite{LS94} directly from MC simulations of the radial distribution function at contact.

In principle, one could also estimate $Z$ only from $c_3$ as $Z=c_3/\eta$ [see Eq.~\eqref{2c.5}]. As shown in Fig.~\ref{mu}(b), the values of $c_3/\eta$ agree very well with those of $1+\frac{1}{3}c_1+\frac{2}{3}c_2+c_3$ up to $\eta\simeq 0.35$, but tend to lie slightly below the latter ones at higher densities. This small discrepancy is just a consequence of the fact that the \emph{exact} function $\beta\mu_0^\ex(\eta,\sigma_0)$ is \emph{not} a cubic polynomial. In fact, as discussed at the end of Sec.\ \ref{sec2}, the only cubic polynomial that is consistent with both Eqs.~\eqref{LSeq} and \eqref{2c.5} is the PY-c/SPT approximation, which is not particularly accurate. Our MD results show that the excess chemical potential $\beta\mu_0^\ex(\eta,\sigma_0)$ can be fitted extremely well by a cubic polynomial for diameters $\sigma_0$ from $\sigma_0=0$ to $\sigma_0\sim\sigma$ (see Fig.~\ref{fig1}). On the other hand, while the choice of the degree of the polynomial is consistent with the exact property $\lim_{\sigma_0\to\infty}\beta\mu_0^\ex(\eta,\sigma_0)/(\sigma_0/\sigma)^3=\text{finite}$, it would be too far-fetched to expect that such an extreme limit coincides with the coefficient $c_3$ fitted in the range $0\leq\sigma_0\lesssim \sigma$. The fact, however, that the coefficient $c_3$ is so close to $\eta Z$ means that the cubic polynomial fitted in the range  $0\leq\sigma_0 \lesssim \sigma$ keeps being a very good approximation even if $\sigma_0\gg \sigma$. Anyway, the recommended route to measure the compressibility factor from a fit of the form \eqref{2.12} for $\sigma_0\lesssim \sigma$ is $Z=1+\frac{1}{3}c_1+\frac{2}{3}c_2+c_3$ rather than $Z=c_3/\eta$.}

{For future reference of researchers interested in the chemical potential of HS fluids,
we present in Table \ref{table3} the numerical values of the MD data plotted in Figs.~\ref{c0_c1}--\ref{mu}.}

\section{Conclusions}
\label{sec5}

To conclude, this work provides new insights into the properties of the chemical potential of
HS fluids and its relation with the equation of state. A third-degree expression in the
test particle diameter for the chemical potential is shown
to reproduce well that for  HSs with the same diameter as those of  the
host fluid, and also for those tracer particles with smaller and, to some extent, larger diameters (not too close to $\eta=0.49$
in the latter case). It is found that the chemical potential predicted by the
the CS  and
the CSK equations is in close agreement with simulation data. However, it is concluded
that although a third-degree polynomial  in tracer particle diameter
is a very good approximation of the chemical potential, this functional form
cannot be exact. It is also shown that the equation of state of the HS fluid can be accurately
obtained from the polynomial fit of the simulation data for the chemical potential.

Originally
implemented on NVT MC simulations, we have applied in this paper the LS technique\cite{LS94}  to MD simulations. In addition, our
implementation differs from that of Ref.~\onlinecite{LS94} in a few aspects.
First, all four coefficients $c_0$--$c_3$ have been fitted, whereas LS forced $c_0$ and $c_1$
to be equal to their exact values and enslaved $c_3$ to $c_2$ by means of Eq.~\eqref{LSeq},
so that in the end only the coefficient $c_2$ was fitted.  Also, they needed to measure
the compressibility factor $Z$ (from the contact value of the radial distribution function)
independently of the insertion probability measurements, whereas in our case $Z$ is just
another output (in addition to $\beta\mu^\ex$) rather than an input. The excellent agreement
between the fitted $c_0$ and $c_1$ with the exact expressions is an \emph{a posteriori}
confirmation of the accuracy of the results reported in this paper.
We have been able to reach reliable statistical results up to $P_0\approx 2\times 10^{-5}$,
which is about $50$ times smaller than the threshold value considered in Ref.~\onlinecite{LS94}.
Furthermore, our study covers a much larger number of densities.

The LS simulation technique is shown to be an
extremely powerful and adaptable tool to obtain the chemical potential
of tracer particles and the equation of state of HS fluids.
It has also been {shown  that} the BMCSL and BCSK formulas for $c_{2}$
and $c_{3}$ are extremely accurate, and not distinguishable from the numerical data.
Therefore it may be concluded that the equation of state of the {monodisperse} HS fluid {can be considered}
for \emph{most} practical \emph{applications} {to be} a solved analytic problem.

{In the extension to HS binary mixtures of the LS method carried out by Baro\v{s}ov\'a \emph{et al.}\cite{BMLS96} the authors fitted their MC simulated values of $P_0(\eta,\sigma_0)$ to a quartic polynomial. On the other hand, we plan to extend our MD implementation to HS mixtures (binary, ternary, or, more generally, polydisperse) by keeping instead a cubic form since the exact condition $\lim_{\sigma_0\to\infty}\beta\mu_0^\ex(\eta,\sigma_0)/\sigma_0^3=\text{finite}$ still holds for mixtures.
According to Eq.~\eqref{2.11}, the coefficient $c_0$ is the same as in the monodisperse system, while the linear coefficient, once multiplied by $M_3/M_1M_2$, is again the exact $c_1$. As {carried out} in the present paper, these two conditions will be used as confidence tests of the simulation results.}

\begin{acknowledgments}
The research of A.S. has been partially supported by the Spanish Government through Grant No.\ FIS2013-42840-P and by the Regional Government of Extremadura (Spain) through Grant No.\ GR15104 (partially financed by ERDF funds).
D.M.H. would like to thank Dr.\ T. Crane (Department of Physics, Royal Holloway, University of London, UK) for helpful software support.

\end{acknowledgments}

\newpage
\appendix
\section{Chemical potential in the small-size limit}
We consider an $N$-particle HS mixture in $d$ dimensions. The packing fraction of the mixture is $\eta=(N/V)v_d M_d$, where $v_d=(\pi/4)^{d/2}/\Gamma(1+d/2)$ is the volume occupied by a sphere of unit diameter. The Boltzmann factor associated with the potential energy $\Phi_N(\mathbf{r}^N)$ of the mixture is
\beq
e^{-\beta\Phi_N(\mathbf{r}^N)}=\prod_{\alpha=1}^{N-1}\prod_{\gamma=\alpha+1}^N \Theta\left(r_{\alpha\gamma}-\sigma_{\ell_\alpha \ell_\gamma}\right),
\eeq
where $\Theta(x)$ is the Heaviside step function, $r_{\alpha\gamma}=|\mathbf{r}_\alpha -\mathbf{r}_\gamma|$ is the relative distance between particles $\alpha$ and $\gamma$, $\ell_\alpha$ denotes the species particle $\alpha$ belongs to, and $\sigma_{ij}=\frac{1}{2}(\sigma_i+\sigma_j)$.

Now we assume that an extra test particle of diameter $\sigma_0$ is inserted into the fluid.
The canonical ensemble expression for the insertion probability is [see Eq.~\eqref{eq:nkzaa2a}]
\bal
P_0(\eta,\sigma_0)=&\left\langle \prod_{\gamma=1}^N\Theta(r_{0\gamma}-\sigma_{0\ell_\gamma})\right\rangle\nn
=&\frac{\int d\mathbf{r}^N \,e^{-\beta\Phi_N(\mathbf{r}^N)}\int d\mathbf{r}_0\,\prod_{\gamma=1}^N\Theta(r_{0\gamma}-\sigma_{0\ell_\gamma})}{V\int d\mathbf{r}^N \,e^{-\beta\Phi_N(\mathbf{r}^N)}}.
\label{A.10}
\eal
In the limit $\sigma_0\to 0$, we can write
\beq
P_0(\eta,\sigma_0)= P_0(\eta,0)+\dot{P}_0(\eta,0)\sigma_0+\mathcal{O}(\sigma_0^2),
\label{A.2}
\eeq
where the dot denotes a derivative with respect to $\sigma_0$.
The first term on the right-hand side of Eq.~\eqref{A.2} is trivial since
\beq
\int d\mathbf{r}_0\,\prod_{\gamma=1}^N\Theta\left(r_{0\gamma}-\frac{\sigma_{\ell_\gamma}}{2}\right)=V(1-\eta).
\label{A.3}
\eeq
This expresses the fact that, for \emph{any} nonoverlapping configuration of $N$ spheres, the available volume for the test point particle is $V(1-\eta)$. Consequently,
\beq
P_0(\eta,0)=1-\eta.
\label{A.4}
\eeq
As for the derivative $\dot{P}_0(\eta,\sigma_0)$, it is given from Eq.~\eqref{A.10} by
\begin{widetext}
\beq
\dot{P}_0(\eta,\sigma_0)=-\frac{1}{2}\sum_{\alpha=1}^N\frac{\int d\mathbf{r}^N \,e^{-\Phi_N(\mathbf{r}^N)}\int d\mathbf{r}_0\,\delta(r_{0\alpha}-\sigma_{0\ell_\alpha})\prod_{\gamma\neq\alpha}\Theta(r_{0\gamma}-\sigma_{0\ell_\gamma})}{V\int d\mathbf{r}^N \,e^{-\beta\Phi_N(\mathbf{r}^N)}}.
\label{A.5}
\eeq
Making $\sigma_0\to 0$ and assuming again a nonoverlapping configuration of the fluid particles, we can write
\beq
\int d\mathbf{r}_0\,\delta\left(r_{0\alpha}-\frac{\sigma_{\ell_\alpha}}{2}\right)\prod_{\gamma\neq\alpha}\Theta\left(r_{0\gamma}-\frac{\sigma_{\ell_\gamma}}{2}\right)=\Omega_d \lim_{\epsilon\to 0} \int_{0}^{\frac{\sigma_{\ell_\alpha}}{2}+\epsilon}dr_{0\alpha}\,r_{0\alpha}^{d-1}\delta\left(r_{0\alpha}-\frac{\sigma_{\ell_\alpha}}{2}\right)=\Omega_d 2^{1-d}\sigma_{\ell_\alpha}^{d-1},
\label{A.6}
\eeq
\end{widetext}
where $\Omega_d=dv_d 2^d$ is the total solid angle. Therefore,
\beq
\dot{P}_0(\eta,0)=-d\eta\frac{M_{d-1}}{M_d}.
\label{A.7}
\eeq
After insertion of Eqs.~\eqref{A.4} and \eqref{A.7}, Eq.~\eqref{A.2} becomes
\beq
P_0(\eta,\sigma_0)=(1-\eta)\left(1-d\frac{\eta}{1-\eta}\frac{M_1M_{d-1}}{M_d}\frac{\sigma_0}{M_1}\right)+\mathcal{O}(\sigma_0^2).
\label{A.8}
\eeq
Finally, from Eq.~\eqref{1.1} we find
\beq
\beta\mu_0^\ex(\eta,\sigma_0)=c_0(\eta)+c_1(\eta)\frac{M_1M_{d-1}}{M_d}\frac{\sigma_0}{M_1}+\mathcal{O}(\sigma_0^2)
\label{A.9}
\eeq
with
\beq
c_0(\eta)=-\ln(1-\eta),\quad c_1(\eta)=d\frac{\eta}{1-\eta}.
\label{A.9.1}
\eeq
Identifying the test particle as a particle of species $i$ (\ie $\sigma_0=\sigma_i$) and
{focusing on} $d=3$, {it can be readily shown} {that Eqs.~\eqref{A.9} and \eqref{A.9.1} reduce to Eq.~\eqref{2.11c} and \eqref{2.6.0}--\eqref{2.6.1}, respectively}.

Equation~\eqref{A.7} can be obtained by a different route.
Imagine a test particle that can (partially) ``penetrate'' inside the fluid particles,
 \ie it has a \emph{nominal} diameter $\sigma_0<0$ so that the  closest distance $\sigma_{0j}$ between the centers of the test particle and a particle of species $j$  is smaller than $\frac{1}{2}\sigma_{j}$. In that case, Eq.~\eqref{A.10} still holds and, in analogy to Eq.~\eqref{A.3},
\beq
\int d\mathbf{r}_0\,\prod_{\gamma=1}^N\Theta(r_{0\gamma}-\sigma_{0\ell_\gamma})=V-\sum_j N_j v_d(2\sigma_{0j})^d.
\label{A.11}
\eeq
Therefore,
\begin{subequations}
\label{A.12&A.13}
\beq
P_0(\eta,\sigma_0<0)=1-\frac{1}{V}\sum_j N_j v_d(\sigma_0+\sigma_j)^d,
\label{A.12}
\eeq
\beq
\dot{P}_0(\eta,\sigma_0<0)=-\frac{d}{V}\sum_j N_j v_d(\sigma_0+\sigma_j)^{d-1}.
\label{A.13}
\eeq
\end{subequations}
Taking the limit $\sigma_0\to 0$, Eqs.~\eqref{A.12&A.13}
reduce to Eqs.~\eqref{A.4} and \eqref{A.7}.
This in turn shows that  both $P_0(\eta,\sigma_0)$ and $\dot{P}_0(\eta,\sigma_0)$ are continuous at $\sigma_0=0$.

\end{document}